\documentstyle[12pt]{article}
\begin{document}
\global\parskip 10pt
\baselineskip 15pt
\parindent 20pt
\newcommand{\be}{\begin{equation}}
\newcommand{\ee}{\end{equation}}
\newcommand{\bea}{\begin{eqnarray}}
\newcommand{\eea}{\end{eqnarray}}
\begin{titlepage}
\hspace*{\fill}
\begin{center}
\vskip -3.0cm
\begin{flushright}
OKHEP 92-001
\end{flushright}
\vskip 4.0cm
{\Large\bf One-Loop Vilkovisky-DeWitt  Counterterms \\
for 2D Gravity plus Scalar Field Theory}\\
\vskip .5in
R. Kantowski\\
and\\
Caren Marzban\\
\vskip .3 truein
University of Oklahoma, Department of Physics and Astronomy,\\
Norman, OK 73019, USA\\
\vskip .2in
\end{center}
\begin{abstract}
The divergent part of the one-loop off-shell effective action is computed
for a single scalar field coupled to the Ricci curvature  of 2D gravity
($c \phi R$), and self interacting by an arbitrary potential term $V(\phi)$.
The Vilkovisky-DeWitt effective action is used to compute gauge-fixing
independent results. In our background field/covariant gauge we find that
the Liouville theory is finite on shell. Off-shell, we find a large class
of renormalizable potentials which include the Liouville potential. We also
find that for backgrounds satisfying $R=0$, the Liouville theory is finite
off shell, as well.

\end{abstract}
\vspace*{\fill}
August 1992\\
e-mail: kantowski or marzban @phyast.nhn.uoknor.edu or @uokniels.bitnet
\end{titlepage}

\newpage

\section{Introduction and Outline}

The subject of 2D gravity has recently been revived, primarily by the use of
the $\sigma$-model representation of string theory. The powerful technology
of string theory and conformal field theory has led to several breakthroughs,
ranging from offering a consistent theory of quantum gravity, to the
derivation of exact non-perturbative results with important conceptual
consequences [1]. Paralleling these advancements there has been
significant progress on the perturbative aspects of the subject as well
[2,3,4].

As a starting point, the classical action that is often taken is one
that reproduces Einstein's field equations $R=0$ in two dimensions [5].
This is done by the introduction of an auxiliary scalar field as a Lagrange
multiplier, which is then elevated to a fully propagating field, due to
the appearance of a kinetic term at the 1-loop level. Having necessitated
the coupling of gravity to a scalar field, one is then faced with a plethora
of candidate classical actions with different couplings and potential terms.
In [3,6] it was shown that, in fact, by field redefinition one can
reduce the number of arbitrary variables in the action to only one - the
potential for the scalar field. In [3], using the machinery of $\sigma$-models,
it was also argued that the Liouville theory defined by an exponential
potential
is renormalizable, and even finite for a special choice of couplings;
this result is well motivated by string theory.

The work of [3] was done in the conformal gauge. Other gauges have been
chosen by various authors [2,4] to study some of the same issues.
The conformal field theory treatments are indeed consistent [2,4,7],
but the perturbative results are plagued with gauge-dependent results
[2], precluding the comparison of the results. This state of affairs
is not unusual, and methods have been devised for eliminating such ambiguities.

If the issues of finiteness or renormalizability
are to be addressed properly, one must be assured of the
existence of a unique, gauge-independent, result. A quantity which in principle
carries all of the information in a quantum field theory is the effective
action, and Vilkovisky and DeWitt [8] have provided us with an algorithm
for obtaining a unique effective action (also called the VD effective action.)
Central in this plan is the introduction of a metric and a connection on
the manifold of fields. The presense of a gauge group $G$ in a theory calls
for a G invariant field metric and the modding out of the G-equivalent orbits
on this space. The connection in a non-gauge invariant theory is given by the
Christoffel symbols of the field metric, but for a gauge theory an additional
non-local (${\cal T}$) term must be appended.
The VD effective action is gauge-independent (as well as reparametrization
invariant); the only caveat is the non-uniqueness of the field metric for
some field theories. Counter terms computed to 1-loop using the VD effective
action amount to the addition of terms (commonly called VD corrections) to
the counter terms computed using the  naive effective action.  As examples of
VD calculations, see [9,10,11]. In this paper we apply the VD
procedure to the theory of 2D-gravity coupled to a scalar field.

An outline of the remainder of this paper is as follows.
In section 2, the classical action is expanded about arbitrary values of
the background fields, and the second-order piece is then modified to
incorporate the 2-dimensional geometric identity $R_{\mu \nu}-(1/2)g_{\mu \nu}
R=0$. In the process of gauge-fixing
the action, in section 3, we motivate the choice of a metric on the field
configuration space. The divergent part of the 1-loop correction to this
gauge-fixed action is also calculated. The divergent contribution of the ghost
fields, at the 1-loop level, is found in section 4. Sections 5 and 6 are
devoted
to the VD corrections - the local (Christoffel), and the nonlocal (${\cal T}$)
contributions, respectively. Finally, in section
7, we discuss some  implications of the combined counter terms.

\section{Perturbations of the classical action}

As we discussed, a variety of 2-dimensional gravity$+$matter theories
can be parametrized by a single potential appearing in the action. For such
theories we wish to calculate the divergent part of the VD effective
action to one-loop by using background field methods [10] (where MTW
conventions
are used). To make the calculation as
simple as we can we use a particular DeWitt gauge that not only makes the
contribution from the non-local
part of the VD connection tractable, but also makes the remainder of the
one-loop operator minimal. The
form of the classical action we start with is:
\be
S=-\int d^2x \sqrt{\bar{g}} \left[ \frac{1}{2} \nabla^{\alpha} \bar{\phi}
  \nabla_{\alpha} \bar{\phi}
+ c\ \bar{\phi}\bar{R} + V(\bar{\phi})\right].
\label{S}
\ee
By expanding $S$ about the background values, $\Phi$ and $g_{\mu \nu}$,
($\bar{\phi}=\Phi +\eta$ and $\bar{g}_{\mu \nu}
=g_{\mu \nu}+h_{\mu \nu}$) the equations of motion for the background
fields are easily read from the terms linear in $\eta$ and $h_{\mu \nu}$,
\bea
\frac{\delta S}{\delta \bar{\phi}(x)}\Big|_{\Phi}& = & -\sqrt{g(x)}
\Biggl[-\Box \Phi+cR+V'(\Phi)\Biggr]_x\;,
\nonumber \\
\frac{\delta S}{\delta \bar{g}_{\alpha \beta}(x)}\Big|_g& = & -\sqrt{g(x)}
\Biggl[-\frac{1}{2}(\nabla^{\alpha}\Phi \nabla^{\beta}\Phi - \frac{1}{2}
g^{\alpha \beta} \nabla^{\gamma}\Phi \nabla_{\gamma} \Phi)
\nonumber\\
&+&c(\nabla^{\alpha} \nabla^{\beta}- g^{\alpha \beta} \Box) \Phi
+\frac{1}{2} g^{\alpha \beta} V(\Phi) \Biggr]_x\;.
\label{deltaS}
\eea
Quadratic terms of the form $\nabla^{\mu} \eta \nabla^{\nu} \hat{h}_{\mu \nu}$
and $\Phi \nabla_{\alpha} \hat{h}^{\alpha \beta} \nabla^{\mu} \hat{h}_{
\beta \mu}$, where $\hat{h}_{\alpha \beta} \equiv h_{\alpha \beta} -
\frac{1}{2}g_{\alpha \beta}h$ is the traceless part of the quantum field
$h_{\alpha \beta}$, appear in the expansion. Such expressions contribute
non-minimal terms to
the second variation of the action, i.e. to the non-gauge fixed one-loop
operator.
Because such non-minimal terms are difficult to deal
with we choose to remove them by e.g.,
an appropriate choice of gauge. However, in 2D there is another way of
removing some of these terms as well, and that is by the use of the
2-dimensional
identity $\bar{R}_{\mu \nu}- \frac{1}{2} \bar{R} \bar{g}_{\mu \nu}=0$.
Expansion of this identity about background yields,
\[
(\Box - R)\hat{h}_{\alpha \beta} = 2P_{\alpha \beta \gamma \delta}
\nabla^{\gamma} \nabla^{\sigma} \hat{h}_{\sigma}^{\delta}\;,
\]
where the trace-free projection tensor $P_{\alpha \beta \gamma \delta}\equiv
\frac{1}{2}
(g_{\alpha \gamma} g_{\beta \delta} + g_{\alpha \delta} g_{\beta \gamma}
- g_{\alpha \beta} g_{\gamma \delta})$.
Upon multiplying by $\Phi$ and $\hat{h}^{\alpha \beta}$, partial inegration
results in the following identity that $\Phi$ and $\hat{h}^{\alpha \beta}$
must satisfy:
\be
\int \sqrt{g}[\frac{1}{2}\Phi \hat{h}^{\alpha \beta}
(\Box-R)\hat{h}_{\alpha \beta}
+ \Phi (\nabla^{\alpha}\hat{h}_{\alpha \gamma})
(\nabla_{\beta}\hat{h}^{\beta \gamma})
+ (\nabla^{\alpha}\Phi)\hat{h}_{\alpha \gamma}
\nabla_{\beta}\hat{h}^{\beta \gamma}]
d^2x = 0\;.
\label{ident}
\ee
Given the two distinct ways of removing the non-minimal terms - via gauge
fixing or via the use of the above identity - we, for the sake of democracy,
do not commit ourselves to either of these ways. Instead, we simply
add the identity (3), weighted
by an arbitrary parameter $\xi$, from the second variation of the action. The
result is
\bea
S^{(2)}\equiv
\frac{1}{2}\phi ^i\frac{\delta ^2S}{\delta \phi ^i\delta \phi ^j}\phi ^j
& = &
- \int d^2x \sqrt{g}
\left\{
\frac{1}{2}g^{\mu \nu}\nabla_{\mu}\eta \nabla_{\nu}\eta
-\nabla_{\mu} \Phi \hat{h}^{\mu \nu}\nabla_{\nu}\eta \right. \nonumber \\
& + & \frac{1}{2}\nabla_{\mu}\Phi \nabla_{\nu}\Phi (\hat{h}_{\alpha}^{\mu}
 \hat{h}^{\alpha \nu}
+\frac{1}{2}h\hat{h}^{\mu \nu}
-\frac{1}{4}g^{\mu \nu}
\hat{h}_{\alpha \beta}\hat{h}^{\alpha \beta}) \nonumber \\
& + & c \left[
-\nabla^{\mu}\eta\nabla^{\nu}\hat{h}_{\mu \nu}
-\frac{1}{2}\eta \Box h
+  \frac{1}{2}\nabla^{\mu}\Phi h\nabla^{\alpha}\hat{h}_{\alpha \mu}
 \right.
\nonumber \\
& + & \frac{1}{4}(1-\xi)\Phi\hat{h}_{\mu \nu}(\Box-R)\hat{h}^{\mu \nu}
 +  \frac{1}{2}(1-\xi)\Phi\nabla_{\alpha}\hat{h}^{\alpha \beta}\nabla^{\mu}
\hat{h}_{\beta \mu}
\nonumber \\
& + &  \frac{1}{2}(2-\xi)\nabla^{\mu}\Phi \hat{h}_{\mu \nu}
\nabla_{\alpha}\hat{h}^{\alpha \nu}
\nonumber \\
& - &
\left.
\frac{1}{2}
\hat{h}_{\alpha}^{\ \sigma}
\hat{h}_{\sigma\beta} \nabla^{\alpha} \nabla^{\beta}\Phi
+\frac{1}{8}(3 \hat{h}_{\alpha\beta}\hat{h}^{\alpha\beta}+h^2)\Box\Phi
\right]
\nonumber\\
& - &
\left.
\frac{1}{4}\hat{h}_{\alpha \beta}\hat{h}^{\alpha \beta}V(\Phi)
+\frac{1}{2}V'(\Phi)\eta h+\frac{1}{2}V''(\Phi)\eta ^2 \right\}\;,
\label{S2}
\eea
where $\phi ^i$ represents the set of quantum fields ${\eta, h_{\alpha
\beta}}$.
The addition of zero, i.e. equation (3), should not affect the final result,
and
as we shall see later, it does not.

\section{Gauge fixing}

A relatively simple choice of quantum gauge is
implemented by the addition of the following to $S^{(2)}$ :
\be
S_{gf}= -\int d^2x  \frac{1}{2}\chi ^{\mu}c_{\mu \nu}\chi ^{\nu} \;,
\ee
with
\bea
c_{\mu \nu}& = &  -\sqrt{g} c \Phi g_{\mu \nu} \delta (x-y)\;,\nonumber \\
\chi^{\nu}& = & \nabla_{\mu}\hat{h}
^{\mu \nu}-\frac{1}{\Phi}\nabla^{\nu}\eta \;.
\eea
This choice of gauge would render $S^{(2)}$ minimal, i.e. all non-minimal terms
would cancel, were it not for the reintroduction
of the non-minimal terms through the identity (3). Below we
will find another gauge which renders $S^{(2)}+S_{gf}$
minimal, where $S^{(2)}$ contains the identity (3) as given in (4).
But first we use gauge (6) to motivate  an essential ingredient in the VD
effective action, a choice of
metric on  the field configuration space.

The quantity $S^{(2)}+
S_{gf}$, with $\xi = 0$,
can be written in the form
\be
- \frac{1}{2} \phi ^i \left[-K_{ij} \Box +L_{ij}^{\mu}\nabla_{\mu}+M_{ij}
\right] \phi^ j \;,
\label{minop}
\ee
which is exactly what is meant by a minimal second order operator (referring to
the
quantity in the brackets). For this
gauge choice $K_{ij}$ reads
\be
K_{ij}=\sqrt{g(x)} \left(
\begin{array}{cc}
1+\frac{c^2}{2\gamma\Phi} & \frac{1}{2}c g^{\gamma \delta} \\
\frac{1}{2}c g^{\alpha \beta} & \gamma \Phi P^{\alpha \beta \gamma \delta}
\end{array}
\right) \delta (x-y)\;,
\label{K}
\ee
where $\gamma \equiv \frac{1}{2}c(\xi-1)$. Equation (\ref{K})
can be used as a metric on the space of fields defined by
$\phi ^i$; however, there is an ambiguity in
defining this metric, for example it could be defined by $S^{(2)}$ alone; in
which case the $\eta \eta$-component would be $1$.
Motivated by this ambiguity,  we will take the configuration space metric to be
 \be
G_{ij}=\sqrt{g(x)} \left(
\begin{array}{cc}
\Theta(\Phi) & \frac{1}{2}c g^{\gamma \delta} \\
\frac{1}{2}c g^{\alpha \beta} & \gamma \Phi P^{\alpha \beta \gamma \delta}
\end{array}
\right) \delta (x-y) \;,
\label{G}
\ee
where $\Theta(\Phi)$ is an arbitrary function of $\Phi$, and $\xi\ne 0$.
The most general local Killing metric on the field space (the gauge-symmetry
group being given
by the diffeomorphisms, see Eqn. \ref{gen})  is similar to this but with $c$
replaced by an arbitrary function of $\Phi$, and
the $hh$-component [$a_1(\Phi)P^{\alpha \beta \mu \nu}(a_2)$] containing two
more arbitrary functions of $\Phi$, where
$P_{\alpha \beta \gamma \delta}(a_2)\equiv \frac{1}{2}
(g_{\alpha \gamma} g_{\beta \delta} + g_{\alpha \delta} g_{\beta \gamma}
- a_2(\Phi) g_{\alpha \beta} g_{\gamma \delta})$. Using this general field
metric in the VD theory introduces non-minimal terms that cannot be removed
by a choice of gauge; this problem may be overcome in a separate article.
Here we concentrate on field metrics of the form (\ref{G}).

For reasons that will become clear (see section 6) we choose to work in
the DeWitt gauge, defined by
\be
\frac{\delta \chi^{\mu}_{DW}}{\delta \phi ^i}=-(c^{-1})^{\mu \nu}
\nabla^{j}_{\nu} G_{ji} \;,
\ee
where $\nabla^j_{\nu}$ are the gauge group generators [see Eqn.(\ref{gen})].
Clearly the DeWitt gauge (assumed linear and homogeneous in the quantum fields)
is determined uniquely by the choice of the
field metric $G_{ij}$ and $c_{\mu\nu}$. For the metric in (\ref{G}) and with
\be
c_{\mu \nu} \equiv 2 \sqrt{g} \gamma \Phi g_{\mu \nu} \delta (x-y),
\label{C}
\ee
the solution is
\be
\chi^{\nu}_{DW} =\nabla_{\mu}\hat{h}^{\mu \nu}
+\frac{\nabla_{\mu}\Phi}{\Phi}\hat{h}^{\mu \nu}
+ \frac{c}{2\gamma\Phi}\nabla^{\nu}\eta
-\frac{\Theta}{2\gamma \Phi}
(\nabla^{\nu} \Phi)\eta
- \frac{c}{4 \gamma \Phi}(\nabla^{\nu} \Phi)h \;.
\label{DWgauge}
\ee
Adopting this gauge results in a minimal gauge-fixed one-loop operator. In this
minimal gauge, $S^{(2)}+S_{gf}$ can again
be written in the form (\ref{minop}), with $K$ given by (\ref{K}), and $L$ and
$M$ given as
follows:
\begin{eqnarray}
L_{\Phi(x) \Phi(y)}^{\mu}
=&&0,
\nonumber\\
L_{\Phi (x) g_{\alpha \beta} (y)}^{\mu} = && -L_{g_{\alpha \beta}(x)
\Phi(y)}^{\mu}
\nonumber \\
=&& \sqrt{g} \left[ (1-\frac{c}{\Phi}-\Theta)P^{\alpha \beta \nu \mu}
\nabla _{\nu}\Phi + \frac{c^2}{4\gamma \Phi} g^{\alpha \beta} \nabla^{\mu} \Phi
 \right]\ \delta (x-y)\;,\label{L} \\
L_{g_{\alpha \beta}(x) g_{\gamma \delta}(y)}^{\mu}
=&&\sqrt{g} \left\{ (\frac{c}{2} + \gamma)\left[P^{\alpha \beta \sigma}_{\ \ \
\ \omega}
P^{\gamma \delta \mu \omega}-P^{\gamma \delta \sigma}_{\ \ \ \omega }
P^{\alpha \beta \mu \omega}\right] \nabla_{\sigma} \Phi
\right\}
\delta (x-y)\;,
\nonumber\\
\;\;\nonumber\\
M_{\Phi(x) \Phi(y)}
=&&
\sqrt{g} \left[V''(\Phi)+\frac{c}{4\gamma}\nabla^{\mu}
[(2\Theta-\frac{c}{\Phi})\frac{\nabla_{\mu}\Phi}{\Phi}]
+\frac{\Theta^2}{2\gamma\Phi}(\nabla \Phi)^2\right] \delta (x-y)\;,
\nonumber\\
M_{\Phi(x) g_{\alpha \beta}(y)} =&& M_{g_{\alpha \beta}(x) \Phi(y)}
\nonumber \\
=&&
\sqrt{g} \Biggl\{
P^{\alpha \beta \gamma \delta}\left[\frac{1}{2}(1+\Theta-\frac{c}{\Phi})
\nabla_{\gamma}\nabla_{\delta}\Phi + \frac{1}{2}\nabla_{\gamma}\Theta
\nabla_{\delta}\Phi \right. \nonumber \\
+&& \left. (-\frac{\Theta}{\Phi}+\frac{c}{2\Phi^2})\nabla_{\gamma}
\Phi \nabla_{\delta}\Phi \right]
\label{M}\\
+&& g^{\alpha \beta} \left[(-\frac{c^2}{8\gamma \Phi^2} + \frac{\Theta
c}{4\gamma\Phi})
(\nabla \Phi)^2 + \frac{c^2}{8\gamma \Phi}\Box \Phi + \frac{1}{2} V'(\Phi)
\right]
\Biggr\}\delta (x-y)\;, \nonumber
\end{eqnarray}

\bea
M_{g_{\alpha \beta(x)} g_{\gamma \delta}(y)}
&&=\sqrt{g} \Biggl\{
P^{\alpha \beta \sigma}_{\ \ \ \ \omega} \left[(1+\frac{2\gamma}{\Phi})
\nabla_{\rho} \Phi \nabla_{\sigma}\Phi - (\gamma +\frac{3}{2}c)
\nabla_{\rho}\nabla_{\sigma}\Phi \right] P^{\gamma \delta \rho \omega}
\nonumber \\
&&+g^{\gamma \delta}\left[(\frac{1}{4}-\frac{c}{2\Phi})
\nabla_{\rho}\Phi \nabla_{\sigma}\Phi \right] P^{\alpha \beta \rho \sigma}
\nonumber \\
&&+P^{\gamma \delta \rho \sigma}\left[(\frac{1}{4}-\frac{c}{2\Phi})
\nabla_{\rho}\Phi \nabla_{\sigma}\Phi \right]g^{\alpha \beta} \nonumber \\
&&+P^{\alpha \beta \gamma \delta} \left[\frac{3}{4}c\Box \Phi
- \frac{1}{4}(\nabla \Phi)^2 - \frac{V}{2} + \gamma \Phi R \right] \nonumber \\
&&+g^{\alpha \beta} g^{\gamma \delta}\left[\frac{c}{4}\Box \Phi
+\frac{c^2}{8 \gamma \Phi}(\nabla \Phi)^2 \right]\nonumber
\Biggr\} \delta(x-y) \;.
\eea

Though we have gone through a labyrinthine path, these equations represent
a simple result. The expression in (\ref{minop}), with $K$,$L$ and $M$ given by
(\ref{K}), (\ref{L}) and (\ref{M}) simply represents $S^{(2)}+S_{gf}$ in a very
specific
gauge, i.e. the DeWitt gauge associated with the field metric (\ref{G}).
The one-loop correction in the naive effective action theory is given by
\bea
W^{(1)}_{S+gf}
&=&\frac{i}{2} \hbox{Tr}\ln \left(G^{ki}[-\frac{1}{2}\left\{K_{ij},\Box\right\}
+\frac{1}{2}
\left\{L^{\mu}_{ij},\nabla_{\mu}\right\}+M_{ij}] \right)\nonumber\\
& = &
\frac{i}{2} \hbox{Tr}\ln \left\{
-G^{km}K_{mi}[\delta^{i}_{j}\Box +L'^{i\mu}_{j}\nabla_{\mu}
+M'^{i}_{j}]
\right\}
 \nonumber \\
& = &
\frac{i}{2} \hbox{Tr}\ln  \left\{ -{G}^{-1}{K}[\hat{1}\Box
+\hat{L}'^{\mu}\nabla_{\mu}
+\hat{M}']
\right\}\;,
\label{trln}
\eea
plus a ghost term. In the above $G^{-1}=G^{km}$ is the inverse of (\ref{G}),
and \footnote[1]{Due to an oversight on our part, only the first terms in these
equations
were included in an original rough draft of this work. We would like to
express our gratitude to F.D. Mazzitelli who found this important error.
Without
his contribution our results would have been in error.}
 $\hat{L}'^{\mu} \equiv -{K}^{-1}L^{\mu}+K^{-1}\nabla^{\mu}K$ and $\hat{M'}
\equiv -{K}^{-1} M -(1/2)K^{-1}(\nabla_{\mu}L^{\mu})+(1/2)K^{-1}(\Box K)$.
The divergent part of (\ref{trln}) is:
\bea
W^{div}_{S+gf}
&=&
\frac{i}{2}\hbox{Tr}\ln \left. [\hat{1}\Box +\hat{L'}^{\mu} \nabla_{\mu}
+ \hat{M'}]\right|_{div}
\nonumber\\
&=&
-\frac{1}{4\pi \epsilon}\int d^2x \sqrt{g}\ \hbox{tr}\ [-\frac{R}{6}\hat{1}
+\frac{1}{4} \hat{L'}^{\mu}\hat{L'}_{\mu} -\hat{M'}] \;,
\label{trlndiv}
\eea
where dimensional regularization ($\epsilon\equiv 2-n$) has been used to
express the irreducible divergence as a simple pole.
{}From (\ref{K}), (\ref{L}), and (\ref{M}) we find
\bea
\hbox{tr}\ \hat{L'}^{\mu}\hat{L'}_{\mu}&=&-\frac{2c(\nabla \Phi)^2}{\gamma
\Phi^2} , \; \nonumber\\
\hbox{tr}\ \hat{M'} & = & (\nabla \Phi)^2\left[\frac{c^2}{4\gamma ^2 \Phi ^2}
+ \frac{c}{2\gamma \Phi ^2} - \frac{2}{\Phi ^2}- \frac{\Theta}{\gamma\Phi}
\right]
\nonumber \\
& + & (\Box \Phi )\left[ \frac{2}{\Phi}+\frac{1}{c}\right]
- \left[ \frac{2V'}{c}-\frac{V}{\gamma \Phi}+2R\right] \;.
\label{trMprime}
\eea
Substitution of these into (\ref{trlndiv}) gives
\bea
W^{div}_{S+gf}
&=&-\frac{1}{4\pi \epsilon} \int d^2x \sqrt{g}
\left\{
\frac{4}{3}R +\frac{2V'}{c} -\frac{V}{\gamma \Phi}
\right.
\nonumber \\
& + &\left. (\nabla \Phi)^2 \left[ -\frac{c^2}{4\gamma ^2 \Phi ^2} -
\frac{c}{\gamma \Phi ^2} + \frac{2}{\Phi ^2}
 + \frac{\Theta}{\gamma\Phi}
\right]
-(\Box \Phi) \left[ \frac{2}{\Phi} + \frac{1}{c} \right]\;
\right\}.
\label{Wdiv}
\eea
This is to be combined with a ghost contribution and a VD correction.
As it stands now the divergent part is clearly gauge and field metric dependent
(the $\gamma$ and $\Theta$ terms).

\section{The ghost contribution}

In this section we shall compute the divergent contribution of the ghost fields
to
the effective action. The ghost operator $\hat{Q}^{-1}$ is defined by
\[(Q^{-1})^{\mu}_{\nu}=
\nabla_{\nu}^{i} \frac{\delta \chi ^{\mu}}{\delta \phi ^i}\;,
\]
which in the DeWitt gauge is simply related to the vertical part of the field
metric $N_{\mu \nu} $,
\bea N_{\mu \nu} &\equiv& \nabla^{i}_{\mu}
G_{ij}\nabla^{j}_{\nu},\nonumber\\
 (Q^{-1})^{\mu}_{\nu}
&=&-(c^{-1})^{\mu\sigma}N_{\sigma\nu}.
\label{N}
\eea
The gauge generators $\nabla_{\alpha}^{\Phi}$ and $\nabla_{\alpha}^{g_{\mu
\nu}}$ are the
generators of infinitesimal coordinate changes of the scalar and the
gravitational
fields, respectively:
\bea
\nabla^{\Phi(y)}_{\mu(x)} & = & (\nabla_{\mu}\Phi)_y \delta(x-y), \nonumber \\
\nabla^{g_{\alpha \beta}(y)}_{\mu(x)} & = & (g_{\mu \alpha} \nabla_{\beta}
+g_{\mu \beta} \nabla_{\alpha})_y \delta(x-y) \;.
\label{gen}
\eea
The ghost contribution to the one-loop effective action is
\be
W^{(1)}_Q=
-i\hbox{Tr}\ln
(\hat{Q}^{-1}),
\ee
and from (\ref{G}), (\ref{N}), (\ref{gen}), and (\ref{C})
we find
\bea
(Q^{-1})^{\alpha(x)}_{\nu(y)} & = & -c^{\alpha \mu}N_{\mu \nu},\nonumber \\
& = & \Biggl\{
\delta^{\alpha}_{\nu}\Box
+\left[\
\delta^{\alpha}_{\nu}
\frac{\nabla^{\sigma}\Phi}{\Phi} -(\frac{c}{\gamma}+2)\frac{1}{\Phi}
g^{\alpha \mu} \nabla_{[\mu}\Phi \delta^{\sigma}_{\nu]}
\right]
\nabla_{\sigma}
\nonumber \\
& + & \frac{R}{2}\delta ^{\alpha}_{\nu}
-\frac{\Theta}{2\gamma\Phi}\nabla^{\alpha}\Phi
\nabla_{\nu}\Phi +\frac{c}{2\gamma \Phi} \nabla^{\alpha}\nabla_{\nu} \Phi
\Biggr\}_x
\delta(x-y)\;.
\label{Q}
\eea
This is a minimal operator, the divergent part of whose Trace-log is
found from (\ref{trlndiv}) to be
\bea
W^{div}_Q
& = &\frac{1}{2\pi \epsilon} \int d^2x \sqrt{g}
\Biggl\{ -\frac{4}{3}R
\nonumber \\
& + &
\left[-\frac{c}{2\gamma}(1+\frac{c}{4\gamma})+\frac{\Theta\Phi}{2\gamma}\right]
\frac{(\nabla \Phi)^2}{\Phi ^2} -\frac{c}{2\gamma} \frac{\Box \Phi}{\Phi}
\Biggr\}.
\label{WQdiv}
\eea

\section{Christoffel symbols}

As discussed in section (1), in order to obtain a unique effective action at
one-loop
Vilkovisky and DeWitt direct us to modify the gauge-fixed quantum action
$S^{(2)}+S_{gf}$  by adding
\be
-\frac12\Gamma^{i}_{j \; k} \phi^j \phi^k S_{,i} \;,
\label{VD}
\ee
where $\Gamma^{i}_{j \; k}$ is the connection on the configuration space
constructed from the metric (\ref{G}) and the gauge (\ref{DWgauge}). Due to the
presence of the equations of motion
$S_{,i} \equiv \frac{\delta S}{\delta \Phi ^i}$ [see Eqn.(\ref{deltaS})]
, this correction
vanishes on-shell, but is frequently non-trivial off-shell. For a gauge theory
the connection is
given in two parts:
\be
\Gamma^{i}_{j \; k} = \left\{ \begin{array}{c} i \\
j \;\; k \end{array} \right\} + {\cal T}^{\ i}_{j\ k} \;,
\ee
where $\phi^i={\eta(x), h_{\mu \nu}(x)}$ in condensed notation. The local
contribution is given by the Christoffel symbols $\{\}$, and the non-local
contribution given by  ${\cal T}$
will be the subject of the next section. For the choice of our metric
(\ref{G}), the Christoffel symbols are given as follows:
\bea
\left\{ \begin{array}{c} \eta(x) \\ \eta(y) \;\; \eta(z) \end{array} \right\}
& = &
-\frac{\Theta}{2c} \delta(y-x) \delta(z-x),
\nonumber\\
\left\{ \begin{array}{c} \eta(x) \\ \eta(y) \;\; h_{\alpha \beta}(z)
        \end{array} \right\}
& = & 0,
\nonumber\\
\left\{ \begin{array}{c} \eta(x) \\ h_{\mu \nu}(y) \;\; h_{\alpha \beta}(z)
        \end{array} \right\}
& = &
-\frac{\gamma \Phi}{2c} P^{\mu \nu \alpha \beta} \delta(y-x) \delta(z-x),
\nonumber\\
\left\{ \begin{array}{c} h_{\mu \nu}(x) \\ \eta(y) \;\; \eta(z)
        \end{array} \right\}
 & = &
\frac{1}{2c}(\frac{\Theta^2}{c}+{\Theta'})g_{\mu \nu} \delta(y-x) \delta(z-x),
\nonumber\\
\left\{ \begin{array}{c} h_{\mu \nu}(x) \\ \eta(y) \;\; h_{\alpha \beta}(z)
        \end{array} \right\}
& = &
\left[\frac{1}{2\Phi}P^{\ \ \alpha \beta}_{\mu \nu} +\frac{\Theta}{4c}
g_{\mu \nu}g^{\alpha \beta} \right] \delta(y-x) \delta(z-x),
\nonumber\\
\left\{ \begin{array}{c} h_{\mu \nu}(x) \\ h_{\alpha \beta}(y)
h_{\gamma \delta}(z) \end{array} \right\}
& = &
\Biggl\{
\frac{1}{4}[P^{\ \ \gamma \delta}_{\mu \nu}g^{\alpha \beta}+
P^{\ \ \alpha \beta}_{\mu \nu}g^{\gamma \delta}]
\nonumber\\
& - & \frac{1}{2}[P_{\mu \nu}^{\ \ \alpha (\gamma}g^{\delta) \beta} +
P_{\mu \nu}^{\ \ \gamma (\alpha}g^{\beta) \delta}] \\
& - & \frac{\gamma}{2c}[(1+\frac{c}{\gamma})-\frac{\Theta \Phi}{c}]
g_{\mu \nu}P^{\alpha \beta \gamma \delta}\Biggr\}
\delta(y-x) \delta(z-x)
\nonumber
\;.
\eea
Because the Christoffel connection symbols are local functions of the
background, as are the equations of motion, they will contribute only to
$M_{ij}$ and
not to $K_{ij}$ or $L^{\mu}_{ij}$ in (\ref{trln}) and hence only through the
trace tr($ M'$) to (\ref{trlndiv}).
The addition  to (\ref{trMprime}) is
\bea
\Delta \hbox{tr} (\hat{M}') & = & - (K^{-1})^{ij}
 \left\{ \begin{array}{c} k \\
j \;\; i \end{array} \right\}
 \frac{\delta S}{\delta \Phi ^k},
\nonumber\\
& = &
\left[\frac{-2\Theta}{c} +\frac{1}{c}
+\frac{1}{\Phi}(\frac{c}{\gamma}+1)\right]\Box \Phi \nonumber \\
& + & \left[\frac{2\Theta}{c^2} - \frac{1}{c \Phi}
(\frac{c}{\gamma}+1)\right]\;V -\frac{V'}{c} -R \;,
\label{deltaMprime}
\eea
which then adds a term
\be
W_{\{\}}^{div}=
\frac{1}{4\pi \epsilon}\int d^2x \sqrt{g}\Delta \hbox{tr} (\hat{M}') \;,
\label{WCdiv}
\ee
to the effective action's divergent part (\ref{Wdiv}).
The remaining non-local (${\cal T}$) contribution is somewhat less straight
forward to evaluate.

\section{The ${\cal T}$ contribution}

To evaluate the non-local (${\cal T}$) part of the connection's contribution
to the effective action (divergent part) we simply make use of the
technique outlined by Barvinsky-Vilkovisky.  This technique relys on having
used the DeWitt gauge, which we have. We do not
present any of the derivation here but only apply it. In general  (${\cal
T}$)'s
contribution to the effective action can be written as a sum of traces of
products of two operators $U_1$ and $U_2$,
(see Sec. 5.8 of [10])
\be
W_{\cal T}^{(1)} =
-\frac{i}{2}(\hbox{Tr}U_1 - \hbox{Tr}U_2)-\frac{i}{4}\hbox{Tr}(U_1)^2
+{\cal O}[(S_{,i})^3]\;,
\label{WT}
\ee
where
\be
U^{\mu}_{1\; \nu}\equiv N^{\mu \beta}\nabla_{\beta}^i({\cal D}_i
\nabla^{k}_{\sigma}) S_{,k} N^{\sigma \gamma}c_{\gamma \nu} \;,
\label{U1}
\ee
\[
U^{\mu}_{2\; \nu}\equiv N^{\mu \beta} S_{,k}
({\cal D}_i \nabla ^{k}_{\beta}){\cal G}^{ij}({\cal D}_j \nabla^{l}_{\sigma})
S_{,l} N^{\sigma \gamma }c_{\gamma \nu} \;,
\]
and where ${\cal G}^{ij}$ is the propagator for the gauge fixed one-loop
operator including the Christoffel term, i.e.  $S+S_{gf}$ plus (\ref{VD}) where
only
the Christoffel connection is used.
The other terms have been introduced before except ${\cal D}_i$ and it is the
field  covariant derivative operator with only the Christoffel symbol as its
connection,
\be
{\cal D}_i \nabla^k_{\sigma(x)} = \frac{\delta
\nabla^j_{\sigma(x)}}{\delta \Phi^i}+\nabla^j_{\sigma(x)}
\left\{ \begin{array}{c} k \\ j \;\; i \end{array} \right\}.
\ee
The $U_1$ and $U_2$ operators act on the space of infinitesimal gauge
transformations and not on the $\phi^i$ field space. Their only contribution to
 the divergent part of (\ref{WT})  in 2D is simply
\be W_{\cal T}^{div}=-\frac{i}{2} \hbox{Tr} U_1\Big|^{div}.
\label{effU1div}
\ee
That this is the case is shown by a dimensional argument\footnote[1]{In this
notation, ${\cal O}[\;]$ signifies the background
field dimension, and it is ${\cal O}[l^0]$ for all of $\gamma$,
$\Phi$, $g_{\mu \nu}$, $c$, and $\Theta$.}:
Divergent contributions come from traces of operators of order ${\cal
O}[l^{-d}]$,
where $d=2$ here. Since
$S_{,i}$ is already of order ${\cal O}[l^{-2}]$ and all other terms are at
least of order ${\cal O}[l^0]$, (\ref{effU1div}) is the only contribution.
Because $U_1$ need only be evaluated to  order ${\cal O}[l^{-2}]$ we can write
[$N^{\mu \nu}\equiv(N_{\mu \nu})^{-1}$, see Eqn.(\ref{Q})],

\[
N^{\mu(x) \nu(y)}=\frac{1}{(\sqrt{g}2\gamma \Phi)_x}g^{\mu \nu}
\frac{1}{\Box_{xy}} +{\cal O}[l^{-1}]\; ,
\]
and hence
\be W_{\cal T}^{div}=\frac{i}{2}\int dxdydz \frac{-1}{\sqrt{g(z)} 2 \gamma
\Phi(z)}
g^{\sigma \mu}(z)\left[\nabla^i_{\mu (z)}({\cal D}_i \nabla^k_{\nu (x)})
S_{,k}\right]\frac{\delta^{\nu}_{\tau}}{\Box
_{xy}}\frac{\delta^{\tau}_{\sigma}}{\Box _{yz}} \Bigg|^{div}\;.
\label{WTdiv1}
\ee
Operators have been freely commuted because their differences contribute $
{\cal O}[l^{-3}]$.
Consequently we can write the quantity in brackets as
\[
\nabla^i_{\mu (z)}({\cal D}_i \nabla^j_{\nu (x)})
S_{,j}\equiv
\left[T_{\mu \nu}^{\ \ \alpha \beta} \nabla_{\alpha} \nabla_{\beta}
\right]_z \delta (z-x) + {\cal O}[l^{-3}]\; ,
\]
thereby defining the tensor $T_{\mu \nu}^{\ \ \alpha \beta} $. Equation
(\ref{WTdiv1}) can
now be evaluated  by use of the following coincidence
limit, derivable from the general algorithms outlined in [10]:
\[
\left. \nabla_{\alpha} \nabla_{\beta}\frac{\hat{1}}{\Box ^2}
\delta (z-w)\right|_
{w \rightarrow z} = -\frac{i}{4 \pi \epsilon} \sqrt{g(z)}g_{\alpha \beta}(z)
\hat{1} \;.
\]
Consequently,
\be
 W_{\cal T}^{div} =  \frac{1}{8 \pi \epsilon} \int d^2z
\sqrt{g}\ \frac{T_{\ \mu\ \alpha}^{\mu\ \alpha}(z)}{2 \gamma \Phi(z)}  \;.
\label{WTdiv2}
\ee
Computation of the trace of $T_{\ \mu\ \alpha}^{\mu\ \alpha}$ is all that
remains.
We note that
since ${\cal O}[S_{,i}]={\cal O}[l^{-2}]$ and ${\cal O}[\nabla^{\Phi}_{\mu}]=
{\cal O}[l^{-1}]$ the
terms ${\cal D}_{\Phi}\nabla^{\Phi}_{\sigma}$ and ${\cal D}_{\Phi}
\nabla^{g_{\mu \nu}}_{\sigma}$ do not contribute to $T$, and only some terms in
${\cal D}_{g_{\mu \nu}}\nabla^{\Phi}_{\sigma}$ and ${\cal D}_{g_{\mu \nu}}
\nabla^{g_{\kappa \lambda}}_{\sigma}$ do. The former can be
 found to be
\[
{\cal D}_{g_{\mu \nu}(w)}\nabla^{\Phi(z)}_{\sigma (x)}=-\delta(z-w)
\left[\frac{\gamma \Phi}{c} P^{\mu \nu , \sigma \tau} \nabla_{\tau}\right]_
{w} \delta(w-x),
\]
and the latter can be cast in the form
\[
{\cal D}_{g_{\mu \nu}(w)}\nabla^{g_{\kappa \lambda}(z)}_{\sigma(x)}
\equiv \delta(z-w) \left[t_{\kappa \lambda\ \ \ , \sigma \tau}^{\ \ \ \mu \nu}
\nabla^{\tau (x)} \right] \delta (w-x) + {\cal O}[l^{-1}] \; ,
\]
with
\begin{eqnarray*}
t_{\kappa \lambda\ \ \ , \sigma \tau}^{\ \ \ \mu \nu}& = &
2\delta^{\mu \nu}_{\sigma (\kappa} g_{\lambda )\tau}
-\delta ^{\mu \nu}_{\kappa \lambda} g_{\sigma \tau} -
[1+\frac{\gamma}{c}-\frac{\Theta \gamma }{c^2}\Phi]g_{\kappa \lambda}
P_{\sigma \tau}^{\ \ \mu \nu} \\
& + & \frac{1}{2}(g_{\sigma \tau}P^{\ \ \mu \nu}_{\kappa \lambda} +
g^{\mu \nu}P_{\kappa \lambda, \sigma \tau}) -
(P^{\ \ \ (\mu}_{\kappa \lambda\sigma}\delta^{\nu)}_{\tau}
+P^{\ \ \ (\mu}_{\kappa \lambda\tau}\delta^{\nu)}_{\sigma})\;.
\end{eqnarray*}
The trace is then given by
\[
T_{\ \mu\ \nu}^{\mu\ \nu} = 2\left[ \frac{\gamma \Phi}{c} P^{\mu \nu}_{\ \ \mu
\nu}
\frac{\delta S}{\delta \Phi} - t^{\ \ \ \mu \nu}_{\kappa \lambda\ \ \ , \mu
\nu}
\frac{\delta S}{\delta g_{\kappa \lambda}} \right] \;,
\]
and by (\ref{WTdiv2}) leads to
\bea
 W_{\cal T}^{div}
& = &
-\frac{1}{8 \pi \epsilon} \int d^2x \sqrt{g} \frac{2}{c}
\Biggl[ \Box \Phi -cR -V'  \nonumber \\
& + & (\frac{1}{\Phi}-\frac{\Theta}{c})(c\Box \Phi -V) \Biggr] \;.
\label{WTdiv}
\eea
Note that this quantity vanishes on shell, as it should and that boundary terms
have been kept.

\section{Conclusions and discussion}

At this point we combine (\ref{Wdiv}) and (\ref{WQdiv})  to get the counter
term as computed using the naive effective action theory,
\bea
W^{div}_N&=&W^{div}_{S+gf}+W^{div}_Q
\label{WNdiv}
\\
&=&-\frac{1}{2\pi \epsilon} \int d^2x \sqrt{g}
\left[2R+\frac{V'}{c}-\frac{1}{2\gamma \Phi}V
\right. \nonumber\\
&+&\left.
\frac{(\nabla \Phi)^2}{\Phi ^2} -\frac{1}{2}(\frac{2}{\Phi}
-\frac{c}{\gamma \Phi}+\frac{1}{c})\Box\Phi\right] \;,
\nonumber
\eea
where no divergent surface terms have been dropped. Note the absence of the
metric term $\Theta$
and the presence of $\gamma$ in this equation. We suspect that the $\gamma$
appears much like gauge parameters appear in the  off-shell naive effective
action through quantum gauge fixing.
Eventhough it appears in the action before gauge fixing through the 2D identity
(\ref{ident}), it also enters in the  gauge fixing term  (\ref{DWgauge}), and,
overall, it disappears on shell. Use of a non-minimal gauge would be required
before the source of the $\gamma$-dependence could be decided.
The $\Theta(\Phi)$ term could conceivably appear because it was in the gauge
fixing term, but it doesn't.

The two VD correction terms (\ref{WCdiv}) and (\ref{WTdiv}) combine to give
\be
\Delta W^{div}_{VD} = W^{div}_{\{\}}+W^{div}_{\cal T} =
-\frac{1}{4\pi\epsilon}\int d^2x \sqrt{g}(\frac{\Theta}{c}
-\frac{c}{\gamma \Phi})(\Box \Phi -\frac{V}{c})\;,
\label{DeltaWVDdiv}
\ee
which contains both $\Theta$ and $\gamma$, and properly vanishes on shell.
Combining (\ref{WNdiv}) and (\ref{DeltaWVDdiv})
yields our final result for the divergent part of the ``unique" 1-loop
effective action:
\be
W_{VD}^{div}=-\frac{1}{2\pi \epsilon} \int d^2x \sqrt{g}
\left[2R+\frac{V'}{c}-\frac{\Theta V}{2c^2}+
\frac{(\nabla \Phi)^2}{\Phi ^2} - \frac{1}{2}(\frac{2}{\Phi}-
\frac{\Theta -1}{c})\Box\Phi\right]\;,
\label{WVDdiv}
\ee
where, again, all divergent surface terms have been kept.

Note that the parameter $\gamma\equiv\frac{1}{2}c(\xi-1)$ appearing in
(\ref{WNdiv}), which represents the 2D identity discussed above, disappears
from the VD-corrected result (\ref{WVDdiv}), as anticipated.
The $\gamma$-dependence in (\ref{WNdiv}) is
in fact consistent with other results in the literature [2,4]. Upon
setting $\gamma=-\frac{1}{2}c$, i.e. $\xi=0$, which amounts to not including
the identity at all, we find these same terms in [2] and [4]. However, the
agreement is not complete; the $(\nabla \Phi)^2$ and the $(\Box \Phi)$ terms
do not agree with those of [2], though they do agree with the revised version
of [4].
Differences may be due to the use of different quantum gauge conditions.
In fact, there
is very little overlap in the choice of gauge conditions used in [2]
and in this calculation, and as a result there are expected discrepancies.
Further, having said that the $\gamma$-dependence in the
$V/(2 \gamma \Phi)$ term agrees with other independent calculations, the
fact that on-shell (i.e. $\Box \Phi =V/c$) the $\gamma$-dependence in
(\ref{WNdiv}) vanishes, lends further support to the correctness of our result.

The VD-corrected effective action (\ref{WVDdiv}) is constructed to circumvent
the
problems of gauge-dependent results alluded to above. Of course, as can be
seen from (\ref{DeltaWVDdiv}), equations (\ref{WNdiv}) and (\ref{WVDdiv})
coincide when on-shell, but
for the off-shell case it is the latter which is designed to be
gauge-independent and unique. ``Uniqueness" refers to quantum gauge fixing
independence, and
it does not in any way rule out the possibility of having results that
depend on the choice of the configuration-space metric. Given an action and a
field metric, one can choose fields to put one or the other in a canonical form
but ordinarily not both.
To standardize both there must be some significant relation between these two
structures. The $\Theta$-dependence in (\ref{WVDdiv}) is a reflection of this
relationship, or lack thereof. Congenially, though,
the $\Theta$-dependence in (\ref{WVDdiv}) does actually vanish on shell, as one
expects.

There are several natural choices of $\Theta$ that one may consider. The
classical action, when expanded about background values of the fields,
contains a $\frac{1}{2}\eta \Box \eta$ term [see Eqn.(\ref{S2})] which suggests
$\Theta =1$
as the $\eta\eta$-component of the field metric. A non-trivial choice
$\Theta=1+{c^2}/({2\gamma \Phi})$,  emerges from the
gauge-fixed action [see (\ref{K})]. Note that this choice itself depends on the
value of the parameter $\gamma$. Clearly  $\Theta=1$ is a preferred choice.

To find out what potentials give rise to on-shell finiteness,
we first partially integrate (\ref{WVDdiv}), keeping the surface terms, to
obtain
\bea
W_{VD}^{div }& = & -\frac{1}{2\pi\epsilon} \int d^2x \sqrt{g}
\Biggl[2R +\frac{V'}{c} -\frac{\Theta V}{2c^2} \nonumber \\
& + & \frac{\Theta -1}{2c}\Box \Phi
- \Box \log \Phi \Biggr] \;.
\label{offshell}
\eea
Substituting the equations of motion $R=({V}/{c}-V')/c$ and
$\Box \Phi =V/c$, the $\Theta$-dependence vanishes, leaving us with
\be
W_{VD}^{div}= -\frac{1}{2\pi\epsilon} \int d^2x \sqrt{g}
\frac{1}{c}\left[\frac{3}{2}\Box \Phi -V'-2c \Box\log\Phi\right] \;,
\label{onshell}
\ee
where all quantities take their on-shell values. The last term, albeit
a total derivative, makes it difficult to evaluate the integrand without
knowledge of the explicit solutions.
Discarding surface terms now allows us to find the potentials for
which the theory is finite on-shell, i.e. for which (\ref{onshell}) vanishes,
modulo
surface terms. Since $V$ itself is a total derivative on shell ($c \Box \Phi
$),
if it satisfies
\[
-V' = \frac{\alpha}{c} V \;,
\]
whose solution is the form
\be
V=\mu  \exp\{-\frac{\alpha \Phi}{c}\} \;,
\label{finite}
\ee
where $\mu$ and $\alpha$ are constants, then the theory is finite.
We conclude that the Liouville theory is finite on shell.

Off-shell, our gauge-fixing independent VD effective action becomes
significant. For example,
we also find potentials for which (\ref{offshell}) is zero off-shell (again
modulo surface terms),
but only for the specific case of  a flat metric ($R=0$).
In this case, with the choice $\Theta =$constant, we again find the potentials
(\ref{finite}). As a
special case, this makes the Liouville theory finite, off-shell as well. It is
worth noting that this off-shell result follows from the VD effective action
(\ref{offshell}) and not from the naive one (\ref{WNdiv}), since the latter
contains the
non-surface term $\Box \Phi/\gamma \Phi$.

The renormalizability of these models depends on the choice of $V(c_i,\Phi)$,
as well, where $c_i$ are coupling constants in $V$.
It is easy to see that the renormalizations
\bea
c & \rightarrow & c\;, \nonumber\\
c_i & \rightarrow & c_i + \Delta c_i \; \nonumber \\
\Phi & \rightarrow & \Phi -\frac{1}{\pi \epsilon c}\;, \nonumber\\
g_{\mu \nu} & \rightarrow & g_{\mu \nu} \left[1+\frac{\Theta}{4\pi \epsilon
c^2}\right],
\label{renorm}
\eea
absorb the infinities in (\ref{offshell}) into the ``wave functions" $\Phi$
and $g_{\mu \nu}$; note that under these redefinitions $\Theta$ drops out.
In this case we find
\[
V' = 2\pi\ \epsilon c \;\Delta c_i \;\frac{\partial V}{\partial c_i}\equiv
-\frac{\partial V }{\partial b}\;,
\]
whose solution is
\be
V=c^{2}{\cal A}(\frac{\Phi -b}{c}),
\label{renormV}
\ee
where ${\cal A}$ is an arbitrary function of the specified argument, and
$b$ is yet another parameter whose renormalization (accompanying \ref{renorm})
is
\[
b \rightarrow b - \frac{1}{2 \pi \epsilon c}\;.
\]
It is clear that the Liouville potential is also in the family of
renormalizable potentials (with the $\mu$ in (\ref{finite}) being renormalized
as $\mu \rightarrow \mu(1-\alpha/2\pi\epsilon c^2))$. What has emerged
that differs from the conformal gauge analysis is that a much wider family of
potentials are renormalizable.

{\bf Acknowledgements}

We have benefited from useful discussions with D. Birmingham, S. Odintsov,
N. Mohammedi, and B. Whiting whom we gratefully acknowledge. F. Mazzitelli and
A. Tseytlin
are thanked for bringing to our attention crucial errors in a rough draft
of this paper. Our gratitude is expressed
to H.T. Cho for pointing out the relevance and suggesting the calculation
of the Vilkovisky-DeWitt corrections. This work was supported
by the Department of Energy, and the Southern Association for High Energy
Physics (SAHEP) funded by the Texas National Research Laboratory Commission
(TNRLC).

\end{document}